\documentclass[sigconf]{acmart}
\usepackage{booktabs} 
\usepackage{multirow}
\usepackage{subcaption}
\usepackage{caption}
\usepackage{setspace}
\usepackage{amsmath}
\usepackage[english]{babel}
\definecolor{mygray}{gray}{0.5}
\usepackage{float}
\usepackage{mathtools}

\copyrightyear{2018} 
\setcopyright{acmcopyright} 
\acmConference[SIGIR '18]{The 41st International ACM SIGIR Conference on Research and Development in Information Retrieval}{July 8--12, 2018}{Ann Arbor, MI, USA}
\acmPrice{15.00}
\acmDOI{10.1145/3209978.3209982}
\acmISBN{978-1-4503-5657-2/18/07}

\begin{document}
\title{Towards Better Text Understanding and Retrieval through Kernel Entity Salience Modeling}

\author{Chenyan Xiong}
\affiliation{
\institution{
Carnegie Mellon University}
}
\email{cx@cs.cmu.edu}

\author{Zhengzhong Liu}
\affiliation{
\institution{
Carnegie Mellon University}
}
\email{liu@cs.cmu.edu}

\author{Jamie Callan}
\affiliation{
\institution{
Carnegie Mellon University}
}
\email{callan@cs.cmu.edu}

\author{Tie-Yan Liu}
\affiliation{
\institution{Microsoft Research}
}
\email{tie-yan.liu@microsoft.com}

\begin{abstract}
This paper presents a Kernel Entity Salience Model (\texttt{KESM}) that improves text understanding and retrieval by better estimating entity salience (importance) in documents.
KESM represents entities by knowledge enriched distributed representations, models the interactions between entities and words by kernels, and combines the kernel scores to estimate entity salience.
The whole model is learned end-to-end using entity salience labels.
The salience model also improves ad hoc search accuracy, providing effective ranking features by modeling the salience of query entities in candidate documents.
Our experiments on two entity salience corpora and two TREC ad hoc search datasets demonstrate the effectiveness of \texttt{KESM} over frequency-based and feature-based methods.
We also provide examples showing how \texttt{KESM} conveys its text understanding ability learned from entity salience to search.
\end{abstract}


\maketitle

\section{Introduction}

Natural language understanding has been a long desired goal in information retrieval. 
In search engines, the process of text understanding begins with the representations of query and documents.
The representations can be bag-of-words, the set of words in the text, or bag-of-entities, which uses automatically linked entity annotations to represent texts~\cite{ESR,Xiong2016BOE,raviv2016document,SELM}.

With the representations, the next step is to estimate the term (word or entity) importance in text, which is also called term \emph{salience} estimation~\cite{dunietz2014new,dojchinovski2016crowdsourced}.
The ability to know which terms are salient (important and central) to the meaning of texts is crucial to many text-related tasks. In ad hoc search, the document ranking is often determined by the salience of query terms in them, which is typically estimated by combining frequency-based signals such as term frequency and inverse document frequency~\cite{croft2010search}.

Effective as it is, frequency is not equal to salience. For example, a Wikipedia article about an entity may not repeat the entity the most frequently; a person's homepage may only mention her name once; a frequently mentioned term may be a stopword. 
In word-based retrieval, many approaches have been developed to better estimate term importance~\cite{blanco2012graph}. However, in entity-based representations~\cite{ESR,xiong2017duet,raviv2016document}, while entities convey richer semantics~\cite{bast2016semantic},
entity salience estimation is a rather immature task~\cite{dunietz2014new, dojchinovski2016crowdsourced} and its effectiveness in search has not yet been explored.

This paper focuses on improving text understanding and retrieval by better estimating \textit{entity salience} in documents. 
We present a Kernel Entity Salience Model (\texttt{KESM}) that estimates entity salience end-to-end using neural networks.
Given annotated entities in a document, 
\texttt{KESM} represents them using Knowledge Enriched Embeddings and models the interactions between entities and words using a Kernel Interaction Model~\cite{K-NRM}.
In the entity salience task~\cite{dunietz2014new}, the kernel scores from the interaction model are combined by \texttt{KESM} to estimate entity salience, and the whole model, including the Knowledge Enriched Embeddings and  Kernel Interaction Model, is learned end-to-end using a large number of salience labels.

\texttt{KESM} also improves ad hoc search by modeling the salience of query entities in candidate documents.
Given a query-document pair and their entities, \texttt{KESM} uses its kernels to model the interactions of \emph{query entities} with the entities and words in the document. It then merges the kernel scores to ranking features and combines these features to rank documents. 
In ad hoc search, \texttt{KESM} can either be trained end-to-end when sufficient ranking labels are available, or be first pre-trained on the salience task and then adapted to search as a salience ranking feature extractor.

Our experiments on a news corpus~\cite{dunietz2014new} and a scientific proceeding corpus~\cite{ESR} demonstrate \texttt{KESM}'s effectiveness in the entity salience task.
It outperforms previous frequency-based and feature-based models by large margins, while requires much less linguistic pre-processing than the feature-based model.
Our analyses find that \texttt{KESM} has a better balance on popular (head) entities and rare (tail) entities when predicting salience. In contrast, frequency-based or feature-based methods are heavily biased towards the most popular entities---less attractive to users as they are more expected.
Also, \texttt{KESM} is less sensitive to document length while frequency-based methods are not as effective on shorter documents.

Our experiments on TREC Web Track search tasks show that \texttt{KESM}'s text understanding ability in estimating entity salience also improves search accuracy. 
The salience ranking features from \texttt{KESM}, pre-trained on the news corpus, outperform both word-based and entity-based features in learning to rank, despite various differences in the salience and search tasks.
Our case studies find interesting examples showing that \texttt{KESM} favors documents centering on query entities over those merely mentioning them.
We find it encouraging that the fine-grained text understanding ability of \texttt{KESM}---the ability to model the consistency and interactions between entities and words in texts---is indeed valuable to ad hoc search.

The next section discusses related work. Section 3 describes the Kernel Entity Salience Model and its application to entity salience estimation.
Section 4 discusses its application to ad hoc search.
Experimental methodology and results for entity salience are presented in Sections 5 and Section 6. 
Those for ad hoc search are in Sections 7 and Section 8. Section 9 concludes.

\section{Related Work}

Representing and understanding texts is a key challenge in information retrieval.
The standard approaches in modern information retrieval represent a text by a bag-of-words; they model term importance using frequency-based signals such as term frequency (TF), inverse document frequency (IDF), and document length~\cite{croft2010search}.
The bag-of-words representation and frequency-based signals are the backbone of modern information retrieval and have been used by many unsupervised and supervised retrieval models~\cite{croft2010search,liu2009learning}.

Nevertheless, bag-of-words and frequency-based statistics only provide shallow text understanding. 
One way to improve the text understanding is to use more meaningful language units than words in text representations.
These approaches include the first generation of search engines that were based on controlled vocabularies~\cite{croft2010search} and also the recent entity-oriented search systems which utilize knowledge graphs in search~\cite{xiong2015fbexpansion,daltonentity,raviv2016document,liu2015latent, ESR}.
In these approaches, texts are often represented by entities, which introduce information from knowledge graphs to search systems.

In both word-based and entity-based text representations, frequency signals such as TF and IDF provide good approximations for the importance or salience of terms (words or entities) in the query or documents. 
However, solely relying on frequency signals limits the search engine's text understanding capability; many approaches have been developed to improve \emph{term importance estimation}.

In the word space, the query term weighting research focuses on modeling the importance of words or phrases in the query. For example, Bendersky et al. use a supervised model to combine the signals from Wikipedia, search log, and external collections to better estimate term importance in verbose queries~\cite{Bendersky2011ParameterizedCW};
Zhao and Callan predict the necessity of query terms using evidence from pseudo relevance feedback~\cite{Zhao2010TermNP};
word embeddings have also been used as features in supervised query term importance prediction~\cite{Zheng2015LearningTR}.
These methods in general leverage extra signals to model how important a term is to capture search intents. They can improve the performance of retrieval models compared to frequency-based term weighting.

The word importance in documents can also be estimated by graph-based approaches~\cite{mihalcea2004textrank,blanco2012graph,rousseau2013graph}.
Instead of using isolated words, the graph-based approaches connect words by co-occurrence or proximity.
Then graph ranking algorithms, for example, PageRank, are used to estimate term importance in a document.
The graph ranking scores reflect the centrality and connectivity of words and are able to improve standard retrieval models~\cite{blanco2012graph,rousseau2013graph}.

In the entity space, modeling term importance is even more crucial.
Unlike word-based representations, the entity-based representations are often automatically constructed and inevitably include noises.
The noisy query entities have been a major bottleneck for entity-oriented search and often required manual cleaning~\cite{liu2015latent,daltonentity, SELM}.
Along this line, a series of approaches have been developed to model the importance of entities in a query, for example, latent-space learning to rank~\cite{EsdRank} and hierarchical ranking models~\cite{xiong2017duet}.
These approaches learn the importance of query entities and the ranking of documents jointly using ranking labels. The features used to describe the entity importance include IR-style features~\cite{EsdRank} and NLP-style features from entity linking~\cite{xiong2017duet}.

Nevertheless, previous research on modeling entity salience mainly focused on query representations, while the entities in document representations are still weighted by frequencies, i.e. in the bag-of-entities model~\cite{ESR, xiong2017duet}.
Recently, Dunietz and Gillick~\cite{dunietz2014new} proposed the entity salience task using the New York Times corpus~\cite{sandhaus2008new};  they consider the entities that are annotated in the expert-written summary to be salient to the article, enabling them to automatically construct millions of training data.
Dojchinovski et al. constructed a deeper study and found that crowdsource workers consider entity salience an intuitive task~\cite{dojchinovski2016crowdsourced}.
Both of them demonstrated that the frequency of an entity is not equal to its salience; a supervised model with linguistic and semantic features is able to outperform frequency significantly, though mixed findings have been found with graph-based methods such as PageRank.

\section{Kernel Entity Salience Model}
\label{sec:model}

\begin{figure*}
\begin{subfigure}{0.48\textwidth}
\includegraphics[width=\textwidth]{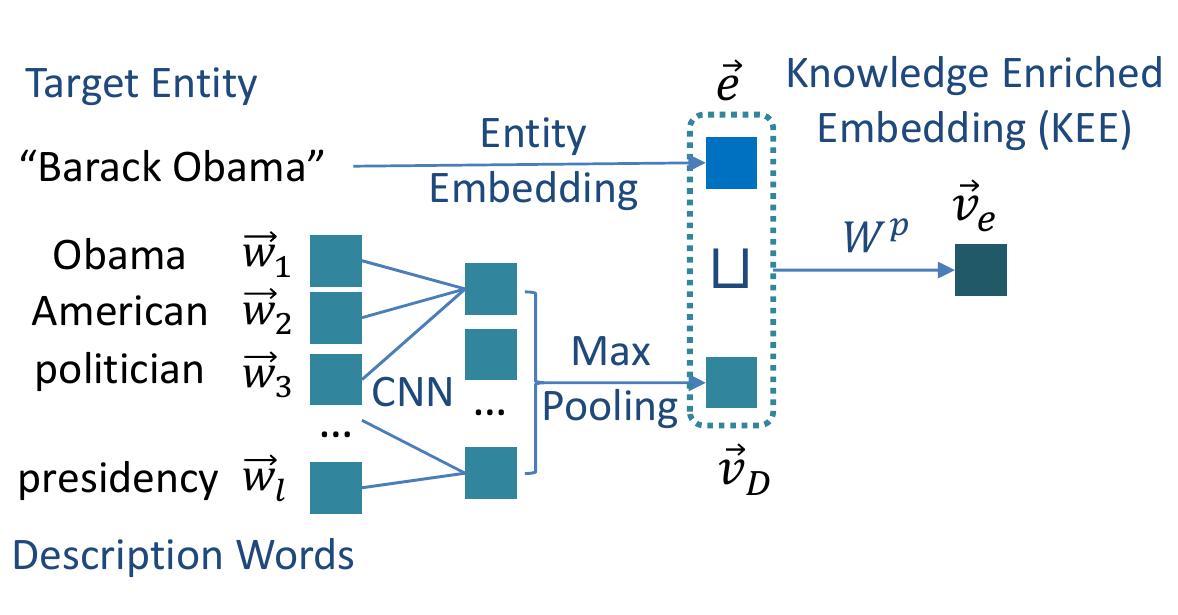}
\caption{Knowledge Enriched Embedding (KEE)\label{fig:kesm_emb}}
\end{subfigure}
\begin{subfigure}{0.48\textwidth}
\includegraphics[width=\textwidth]{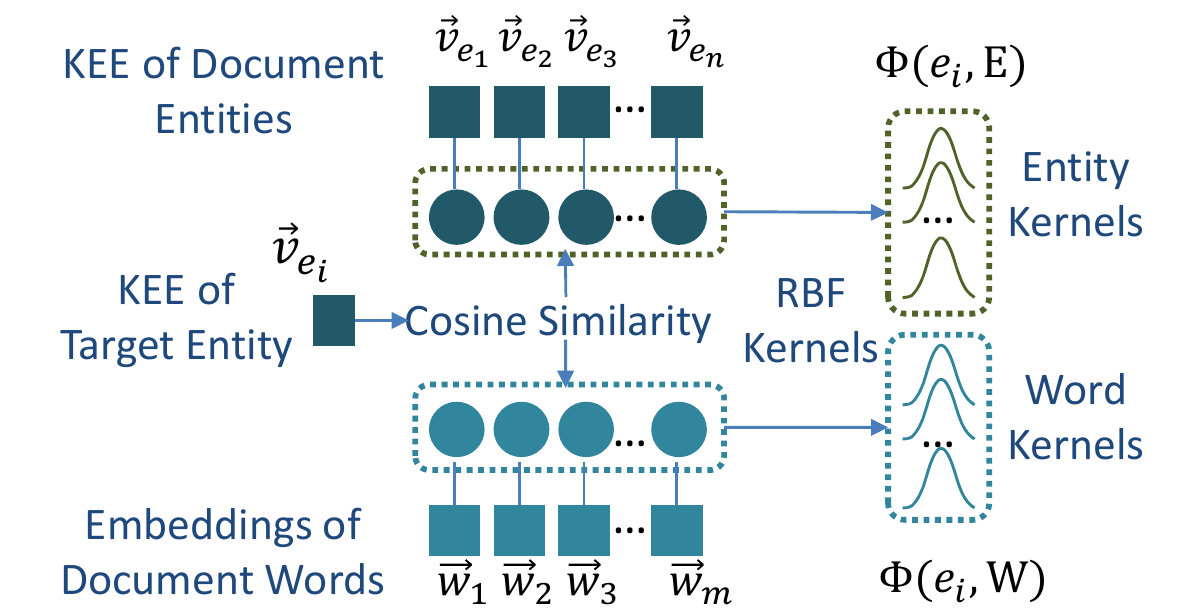}
\caption{Kernel Interaction Model (KIM)\label{fig:kesm_salience}}
\end{subfigure}
\caption{\texttt{KESM} Architecture. (a): Entities are represented using embeddings enriched by their descriptions. (b): The salience of an entity in a document is estimated by kernels that model its interactions with entities and words in the document.
Squares are continuous vectors (embeddings) and circles are scalars (cosine similarities).
\label{fig:kesm}
}
\end{figure*}

This section presents our Kernel Entity Salience Model (\texttt{KESM}).
Compared to the feature-based salience models~\cite{dunietz2014new,dojchinovski2016crowdsourced}, \texttt{KESM} uses neural networks to learn the representation of entities and their interactions for salience estimation. 

The rest of this section first describes the overall architecture of \texttt{KESM} and then how it is applied to the entity salience task.

\subsection{Model Architecture}
As shown in Figure~\ref{fig:kesm}, \texttt{KESM} includes two main components:
the \emph{Knowledge Enriched Embedding} (Figure~\ref{fig:kesm_emb}) and the \emph{Kernel Interaction Model} (Figure~\ref{fig:kesm_salience}).

\textbf{Knowledge Enriched Embedding} (\texttt{KEE}) encodes each entity $e$ into its distributed representation $\vec{v}_e$. It is achieved by first using an embedding layer that maps the entity to an embedding:
\begin{align*}
e &\xrightarrow{V} \vec{e}.  &\text{Entity Embedding}
\end{align*}
$V$ is the parameters of the embedding layer to be learned.

An advantage of entities is that they are associated with external semantics in the knowledge graph, for example, synonyms, descriptions, types, and relations.
Instead of only using $\vec{e}$, \texttt{KEE} enriches the entity representation with its description, for example, the first paragraph of its Wikipedia page.

Specifically, given the description $\mathbb{D}$ of the entity $e$, \texttt{KEE} uses a Convolutional Neural Network (CNN) to compose the words in $\mathbb{D}$: $\{w_1,...,w_p,...,w_l\}$, into one embedding:
\begin{align*}
w_p &\xrightarrow{V} \vec{w}_p,  & \text{Word Embedding}\\
C_p &= W^c \cdot \vec{w}_{p:p+h},  & \text{CNN Filter}\\
\vec{v}_{\mathbb{D}} &= \max (C_1,...,C_p,...,C_{l-h}).  & \text{Description Embedding}
\end{align*}
It embeds the words into $\vec{w}$ using the embedding layer, 
composes the word embeddings using CNN filters, and generates the description embeddings $\vec{v}_{\mathbb{D}}$ using max-pooling.
$W^c$ and $h$ are the weights and length of the CNN.

$\vec{v}_D$ is then combined with the entity embedding $\vec{e}$ by projection:
\begin{align*}
\vec{v}_e &= W^p \cdot (\vec{e} \sqcup \vec{v}_{\mathbb{D}}).
& \text{KEE Embedding}
\end{align*}
$\sqcup$ is the concatenation operator and $W^p$ is the projection weights.
$\vec{v}_e$ is the \texttt{KEE} vector for $e$.
It incorporates the external information from the knowledge graph and is to be learned as part of \texttt{KESM}.


\textbf{Kernel Interaction Model} 
(\texttt{KIM}) models the interactions of a target entity with entities and words in the document using their distributed representations.

Given a document $d$, its annotated entities $\mathbb{E}=\{e_1,...e_i...,e_n\}$, and its words $\mathbb{W}=\{w_1,...w_j...,w_m\}$, \texttt{KIM} models the interactions of a target entity $e_i$  with $\mathbb{E}$ and $\mathbb{W}$ using kernels~\cite{K-NRM, dai2018convolutional}:
\begin{align}
{KIM}(e_i, d) &= \Phi(e_i, \mathbb{E}) \sqcup \Phi(e_i, \mathbb{W}). \label{eq:kim} 
\end{align}

The entity kernels $\Phi(e_i, \mathbb{E})$ model the interaction between $e_i$ and document entities $\mathbb{E}$:
\begin{align}
\Phi(e_i, \mathbb{E}) &= \{\phi_1(e_i, \mathbb{E}),...\phi_k(e_i, \mathbb{E})...,\phi_K(e_i, \mathbb{E}) \},  \\ 
\phi_k(e_i, \mathbb{E}) &= \sum_{e_j \in \mathbb{E}}\exp\left(-\frac{\left(cos(\vec{v}_{e_i}, \vec{v}_{e_j}) - \mu_k\right)^2}{2 \sigma_k^2}\right). 
\end{align}
$\vec{v}_{e_i}$ and $\vec{v}_{e_j}$ are the KEE embeddings of $e_i$ and $e_j$.
 $\phi_k(e_i, \mathbb{E})$ is the $k$-th RBF kernel with mean $\mu_k$ and variance $\sigma_k^2$. If $(\mu_k=1, \sigma_k \rightarrow \infty)$, $\phi_k$ counts the entity frequency. Otherwise, it models the interactions between the target entity $e_i$ and other entities in the \texttt{KEE} representation space. 
One view of kernels is that they count the number of entities whose similarities with $e_i$ are in its region ($\mu_k, \sigma_k^2$); the other view is that the kernel scores are the votes from other entities in a certain neighborhood (kernel region) of the current entity.

Similarly, the word kernels $\Phi(e_i, \mathbb{W})$ model the interactions between $e_i$ and document words $\mathbb{W}$:
\begin{align}
\Phi(e_i, \mathbb{W}) &= \{\phi_1(e_i, \mathbb{W}),...\phi_k(e_i, \mathbb{W})...,\phi_K(e_i, \mathbb{W}) \}, \\
\phi_k(e_i, \mathbb{W})&= \sum_{w_j \in \mathbb{W}} \exp\left(-\frac{\left(cos(\vec{v}_{e_i}, \vec{w}_j) - \mu_k\right)^2}{2 \sigma_k^2}\right). \label{eq:ew_kernel}
\end{align}
$\vec{w}_j$ is the word embedding of $w_j$, mapped by the same embedding parameters ($V$).
The word kernels $\phi_k(e_i, \mathbb{W})$ model the interactions between $e_i$ and document words, gathering `votes' from words for $e_i$ in the corresponding kernel regions.

For each entity $e_i$, \texttt{KEE} encodes it to $\vec{v}_{e_i}$ and \texttt{KIM} models its interactions with entities and words in the document.
The kernel scores $KIM(e_i, d)$ include signals from three sources: the description of the entity in the knowledge graph, its interactions with the document entities, and its interactions with the document words.
The utilization of these kernel scores depends on the specific task: entity salience estimation (Section~\ref{sec:ese}) or document ranking (Section~\ref{sec:rank}).

\subsection{Entity Salience Estimation}
\label{sec:ese}
The application of \texttt{KESM} in the entity salience task is simple. Combining the \texttt{KIM} kernel scores gives the salience score of the corresponding entity: 
\begin{align}
f(e_i, d) &= W^s \cdot KIM(e_i, d) + b^s. \label{eq:kernel_weights}
\end{align}
$f(e_i, d)$ is the salience score of $e_i$ in $d$. $W^s$ and $b^s$ are parameters for salience estimation.

\textbf{Learning:} The entity salience training data are labels about document-entity pairs that indicate whether the entity is salient to the document.
The salience label of entity $e_i$ to document $d$ is: 
\begin{align*}
y(e_i, d) &= \begin{cases}
+1, & \text{if $e_i$ is a salient entity in } d; \\
-1, & \text{otherwise.}
\end{cases}
\end{align*}

We use pairwise learning to rank~\cite{liu2009learning} to train \texttt{KESM}:
\begin{align}
\sum_{e^+, e^- \in d}   \max(0,  1 - f(e^+, d) + f(e^-, d)), \label{eq:salience_loss}\\
\text{w.r.t.}   \text{  } y(e^+, d) = +1 \text{ }\& \text{ } y(e^-, d) = -1. \nonumber
\end{align}
The loss function enforces \texttt{KESM} to rank the salient entities $e^+$ ahead of the non-salient ones $e^-$ within the same document.

In the entity salience task, \texttt{KESM} is trained end-to-end by back-propagation. 
During training, the gradients from the labels are first propagated to the Kernel Interaction Model (\texttt{KIM}) and then the Knowledge Enriched Embedding (\texttt{KEE}).
\texttt{KESM} updates the kernel weights;
\texttt{KIM} converts the gradients from kernels to `expectations' on the distributed representations---how the entities and words should be allocated in the space to better reflect salience;  \texttt{KEE} updates its embeddings and parameters according to these `expectations'.
The knowledge learned from the training labels is encoded and stored in the model parameters, mainly the embeddings~\cite{K-NRM}.

\begin{figure}
\includegraphics[width=0.9\columnwidth]{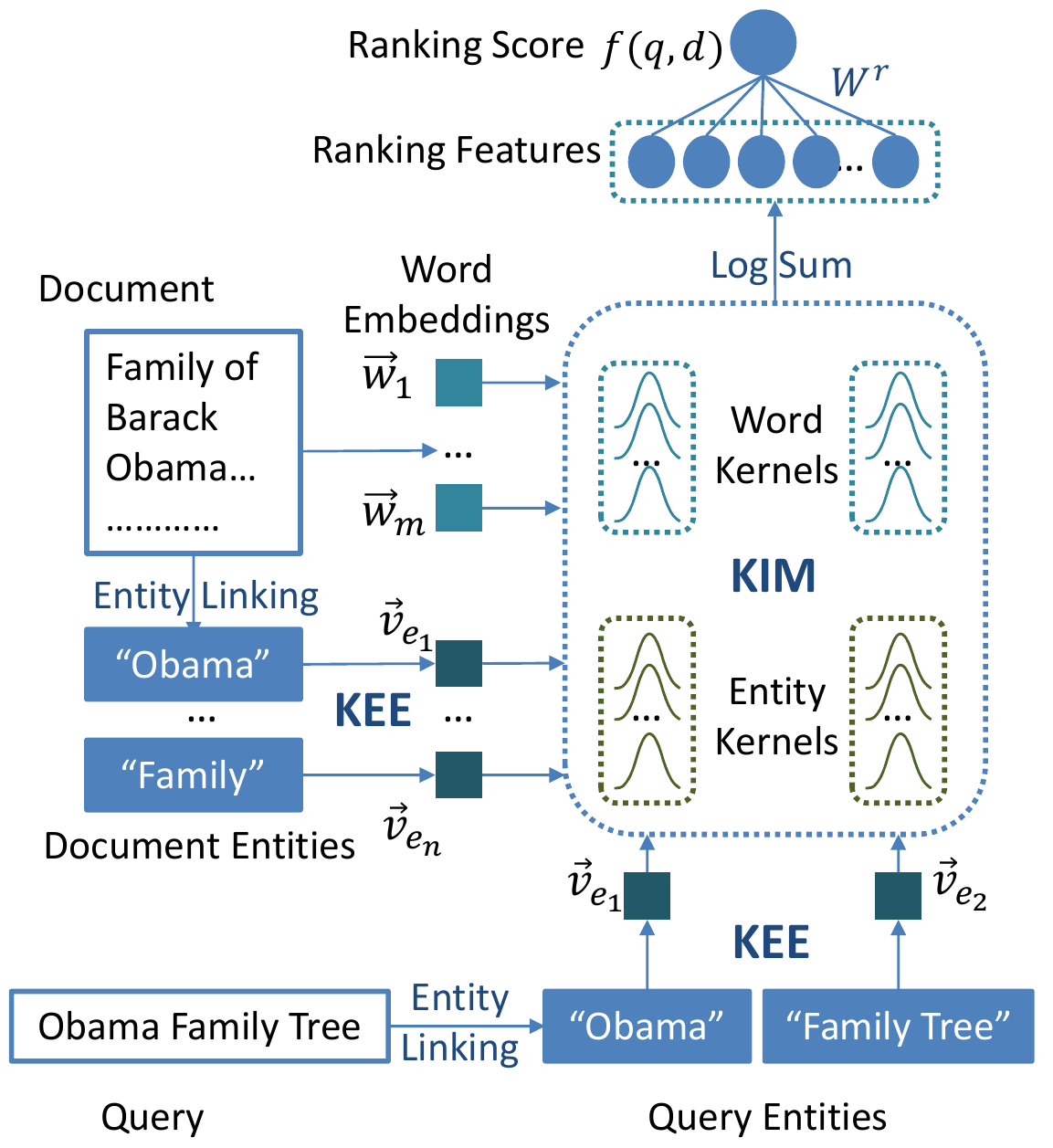}
\caption{Ranking with KESM. \texttt{KEE} embeds the entities. \texttt{KIM} calculates the kernel scores of query entities VS. document entities and words. 
The kernel scores are combined to ranking features and then to the ranking score.
\label{fig:kesm_rank}
}
\end{figure}

\section{Ranking with Entity Salience}
\label{sec:rank}

This section presents the application of \texttt{KESM} in ad hoc search.

\textbf{Ranking:} 
Knowing which entities are salient in a document indicates a deeper text understanding ability~\cite{dunietz2014new,dojchinovski2016crowdsourced}. 
The improved text understanding should also improve search accuracy: the salience of query entities in a document reflects how focused the document is on the query, which is a strong indicator of relevancy.
For example, a web page that exclusively discusses Barack Obama's family is more relevant to the query ``Obama Family Tree'' than those that just mention his family members.

The ranking process of \texttt{KESM} following this intuition is illustrated in Figure~\ref{fig:kesm_rank}.
It first calculates the kernel scores of the query entities in the document using \texttt{KEE} and \texttt{KIM}. 
Then it merges the kernel scores from multiple query entities to ranking features and uses a ranking model to combine these features.

Specifically, given query $q$, query entities $\mathbb{E}^q$, candidate document $d$, document entities $\mathbb{E}^d$, and document words $\mathbb{W}^d$, the ranking score is calculated as:
\begin{align}
f(q, d) &= W^r \cdot \Psi(q, d), \label{eq:kesm_rank} \\
\Psi(q, d) &= \sum_{e_i \in \mathbb{E}^q} \log\left(\frac{KIM(e_i, d)}{|\mathbb{E}^d|}\right). \label{eq:norm}
\end{align}
$KIM(e_i, d)$ are the kernel scores of the query entity $e_i$ in document $d$, calculated by the \texttt{KIM} and \texttt{KEE} modules described in last section.
$|\mathbb{E}^d|$ is the number of entities in $d$.
$W^r$ is the ranking parameters and $\Psi(q, d)$ are the salience ranking features.

Several adaptations have been made to apply \texttt{KESM} in search.
First, Equation~(\ref{eq:norm}) normalizes the kernel scores by the number of entities in the document ($|\mathbb{E}^d|$), making them more comparable across \emph{different} documents. In the entity salience task, this is not required because the goal is to distinguish salient entities from non-salient ones in the \emph{same} document.
Second, there can be multiple entities in the query and their kernel scores need to be combined to model query-document relevance. 
The combination is done by log-sum, following language model approaches~\cite{croft2010search}.

\begin{table}[t]
\caption{Datasets used in the entity salience task. New York Times are news articles and salient entities are those in the expert-written news summaries. Semantic Scholar are paper abstracts and salient entities are those in the titles.
\label{tab:dataset}
}
\begin{tabular}{l|ccc|ccc}
\hline \hline
& \multicolumn{3}{c|}{\textbf{New York Times}}  & \multicolumn{3}{c}{\textbf{Semantic Scholar}}\\ \hline
& Train  & Dev   & Test & Train  & Dev   & Test \\ \hline
\# of Documents &  
526k & 64k & 64k &
800k & 100k & 100k
\\ \hline
Entities Per Doc &   
198 & 197 & 198 &
66 & 66 & 66
\\ \hline
Salience Per Doc &   
27.8 & 27.8 & 28.2 &
7.3 & 7.3 & 7.3
\\ \hline
Unique Word &
609k & 278k & 281k &
921k & 300k & 301k
\\ \hline
Unique Entity &
622k & 319k & 317k &
331k & 162k & 162k
\\ \hline 
\end{tabular}

\end{table}

\begin{table*}[t]
\caption{Entity salience features used by the LeToR baseline~\cite{dunietz2014new}. The features are extracted via various natural language processing techniques, as listed in the \textbf{Source} column.
\label{tab:feature}
}
\begin{tabular}{l|l|l}
\hline \hline
\textbf{Name} & \textbf{Description} & \textbf{Source} \\ \hline
Frequency & The frequency of the entity & Entity Linking\\ 
First Location & The location of the first sentence that contains the entity & Entity Linking \\ \hline
Head Word Count & The frequency of the entity's first head word in parsing & Dependency Parsing \\
Is Named Entity & Whether the entity is considered as a named entity & Named Entity Recognition \\
Coreference Count & The coreference frequency of the entity's mentions & Entity Coreference Resolution \\  \hline
Embedding Vote & Votes from other entities through cosine embedding similarity & Entity Embedding (Skip-gram) \\
\hline \hline
\end{tabular}
\end{table*}

\textbf{Learning:}
In the search task, \texttt{KESM} is trained using standard pairwise learning to rank and relevance labels:
\begin{align}
\sum_{d^+ \in D^+, d^- \in D^-} \max(0, 1 - {f}(q, d^+) + {f}(q, d^-)). \label{eq:ltr}
\end{align}
$D^+$ and $D^-$ are the relevant and irrelevant documents.  $f(q, d^+)$ and $f(q, d^-)$ are the ranking scores calculated by Equation~(\ref{eq:kesm_rank}). 

There are two ways to train \texttt{KESM} for ad hoc search. First, when sufficient ranking labels are available, for example, in commercial search engines, the whole \texttt{KESM} model can be learned end-to-end by back-propagation from Equation~(\ref{eq:ltr}).
On the other hand, when not enough ranking labels are available for end-to-end learning, the \texttt{KEE} and \texttt{KIM} can be first trained using the labels from the entity salience task. Only the ranking parameters $W^r$ need to be learned from relevance labels.
As a result, the knowledge learned from the salience labels is adapted to ad hoc search through the ranking features,
which can be used in any learning to rank system.

\section{Experimental Methodology for Entity Salience Estimation}
This section presents the experimental methodology for the entity salience task.
It mainly follows the setup by Dunietz and Gillick~\cite{dunietz2014new} with some revisions to facilitate the applications in search. An additional dataset is also introduced.

\textbf{Datasets}\footnote{Available at \url{http://boston.lti.cs.cmu.edu/appendices/SIGIR2018-KESM/}} used include \emph{New York Times} and \emph{Semantic Scholar}.

The \emph{New York Times} corpus has been used in previous work~\cite{dunietz2014new}. It includes more than half million news articles and expert-written summarizes~\cite{sandhaus2008new}. 
Among all entities annotated on a news article, those that also appear in the summary of the article are considered as salient entities; others are not~\cite{dunietz2014new}. 

The \emph{Semantic Scholar} corpus contains one million randomly sampled scientific publications from the index of 
\url{SemanticScholar.org}, the academic search engine from Allen Institute for Artificial Intelligence. The full texts of the papers are not released. Only the abstract and title of the paper content are available. We treat the entities annotated on the abstract as the candidate entities of a paper and those also annotated on the title as salient.

The entity annotations on both corpora are Freebase entities linked by TagMe~\cite{TagMe}. 
\emph{All annotations} are included to ensure coverage, which is important for effective text representations~\cite{ESR, raviv2016document}.

The statistics of the two corpora are listed in Table~\ref{tab:dataset}. The Semantic Scholar corpus has shorter documents (paper abstracts) and a smaller entity vocabulary because its papers are mostly in the computer science and medical science domains.

\begin{table*}[t]
\centering
\caption{
Entity salience performances on New York Times and Semantic Scholar.
(E), (W), and (K) mark the resources used by \texttt{KESM}: Entity kernels, Word kernels, and Knowledge enrichment.  \texttt{KESM} is the full model.
Relative performances over \texttt{LeToR} are shown in the percentages. W/T/L are the number of documents a method improves, does not change, and hurts, compared to \texttt{LeToR}. $\dagger$, $\ddagger$, $\mathsection$, and $\mathparagraph$ mark the statistically significant improvements over \texttt{Frequency}$^\dagger$, \texttt{PageRank}$^\ddagger$, \texttt{LeToR}$^\mathsection$, and \texttt{KESM (E)}$^\mathparagraph$.
\label{tab:overall_salience}
}

\begin{tabular}{l|lr|lr|lr|lr|c}
\hline \hline
\multicolumn{10}{c}{\textbf{New York Times}} \\ \hline
\bf{Method}& \multicolumn{2}{c|}{\bf{Precision@1}}& \multicolumn{2}{c|}{\bf{Precision@5}}& \multicolumn{2}{c|}{\bf{Recall@1}}& \multicolumn{2}{c|}{\bf{Recall@5}}&\bf{W/T/L}\\ \hline
\texttt{Frequency}
 & ${0.5840}$ & $ -8.53\%  $

 & ${0.4065}$ & $ -11.82\%  $

 & ${0.0781}$ & $ -11.92\%  $

 & ${0.2436}$ & $ -14.44\%  $

& 5,622/38,813/19,154

\\
\texttt{PageRank}
 & ${0.5845}^{\dagger }$ & $ -8.46\%  $

 & ${0.4069}^{\dagger }$ & $ -11.73\%  $

 & ${0.0782}^{\dagger }$ & $ -11.80\%  $

 & ${0.2440}^{\dagger }$ & $ -14.31\%  $

& 5,655/38,841/19,093

\\ \hline
\texttt{LeToR}
 & $0.6385$ & --  & $0.4610$ & --  & $0.0886$ & --  & $0.2848$ & --  & --/--/--\\
 \hline
\texttt{KESM (E)}
 & ${0.6470}^{\dagger \ddagger \mathsection }$ & $ +1.33\%  $

 & ${0.4782}^{\dagger \ddagger \mathsection }$ & $ +3.73\%  $

 & ${0.0922}^{\dagger \ddagger \mathsection }$ & $ +4.03\%  $

 & ${0.3049}^{\dagger \ddagger \mathsection }$ & $ +7.05\%  $

& 19,778/27,983/15,828

\\
\texttt{KESM (EK)}
 & ${0.6528}^{\dagger \ddagger \mathsection \mathparagraph }$ & $ +2.24\%  $

 & ${0.4769}^{\dagger \ddagger \mathsection }$ & $ +3.46\%  $

 & ${0.0920}^{\dagger \ddagger \mathsection }$ & $ +3.82\%  $

 & ${0.3026}^{\dagger \ddagger \mathsection }$ & $ +6.27\%  $

& 18,619/29,973/14,997

\\
\texttt{KESM (EW)}
 & ${0.6767}^{\dagger \ddagger \mathsection \mathparagraph }$ & $ +5.98\%  $

 & ${0.5018}^{\dagger \ddagger \mathsection \mathparagraph }$ & $ +8.86\%  $

 & ${0.0989}^{\dagger \ddagger \mathsection \mathparagraph }$ & $ +11.57\%  $

 & ${0.3277}^{\dagger \ddagger \mathsection \mathparagraph }$ & $ +15.08\%  $

& 22,805/26,436/14,348

\\ \hline

\texttt{KESM}
 & $\bf{0.6866}^{\dagger \ddagger \mathsection \mathparagraph }$ & $ +7.53\%  $

 & $\bf{0.5080}^{\dagger \ddagger \mathsection \mathparagraph }$ & $ +10.21\%  $

 & $\bf{0.1010}^{\dagger \ddagger \mathsection \mathparagraph }$ & $ +13.93\%  $

 & $\bf{0.3335}^{\dagger \ddagger \mathsection \mathparagraph }$ & $ +17.10\%  $

& 23,290/26,883/13,416

\\ \hline \hline

\end{tabular}

\vspace{0.2cm}

\begin{tabular}{l|lr|lr|lr|lr|c}
\hline \hline
\multicolumn{10}{c}{\textbf{Semantic Scholar}} \\ \hline
\bf{Method}& \multicolumn{2}{c|}{\bf{Precision@1}}& \multicolumn{2}{c|}{\bf{Precision@5}}& \multicolumn{2}{c|}{\bf{Recall@1}}& \multicolumn{2}{c|}{\bf{Recall@5}}&\bf{W/T/L}\\ \hline

\texttt{Frequency}
& ${0.3944}$ & $ -9.99\%  $

 & ${0.2560}$ & $ -11.38\%  $

 & ${0.1140}$ & $ -12.23\%  $

 & ${0.3462}$ & $ -13.67\%  $

& 11,155/64,455/24,390
 \\
\texttt{PageRank}
& ${0.3946}^{\dagger }$ & $ -9.94\%  $

 & ${0.2561}^{\dagger }$ & $ -11.34\%  $

 & ${0.1141}^{\dagger }$ & $ -12.11\%  $

 & ${0.3466}^{\dagger }$ & $ -13.57\%  $

& 11,200/64,418/24,382

\\ \hline

\texttt{LeToR}
& $0.4382$ & --  & $0.2889$ & --  & $0.1299$ & --  & $0.4010$ & --  & --/--/-- \\ \hline

\texttt{KESM (E)}
& ${0.4793}^{\dagger \ddagger \mathsection }$ & $ +9.38\%  $

 & ${0.3192}^{\dagger \ddagger \mathsection }$ & $ +10.51\%  $

 & ${0.1432}^{\dagger \ddagger \mathsection }$ & $ +10.26\%  $

 & ${0.4462}^{\dagger \ddagger \mathsection }$ & $ +11.27\%  $

& 27,735/56,402/15,863

\\
\texttt{KESM (EK)}
& ${0.4901}^{\dagger \ddagger \mathsection \mathparagraph }$ & $ +11.84\%  $

 & ${0.3161}^{\dagger \ddagger \mathsection }$ & $ +9.43\%  $

 & ${0.1492}^{\dagger \ddagger \mathsection \mathparagraph }$ & $ +14.91\%  $

 & ${0.4449}^{\dagger \ddagger \mathsection }$ & $ +10.95\%  $

& 28,191/54,084/17,725

\\
\texttt{KESM (EW)}
 & ${0.5097}^{\dagger \ddagger \mathsection \mathparagraph }$ & $ +16.31\%  $

 & ${0.3311}^{\dagger \ddagger \mathsection \mathparagraph }$ & $ +14.63\%  $

 & ${0.1555}^{\dagger \ddagger \mathsection \mathparagraph }$ & $ +19.77\%  $

 & ${0.4671}^{\dagger \ddagger \mathsection \mathparagraph }$ & $ +16.50\%  $

& 32,592/50,428/16,980

\\ \hline

\texttt{KESM}
 & $\bf{0.5169}^{\dagger \ddagger \mathsection \mathparagraph }$ & $ +17.96\%  $

 & $\bf{0.3336}^{\dagger \ddagger \mathsection \mathparagraph }$ & $ +15.47\%  $

 & $\bf{0.1585}^{\dagger \ddagger \mathsection \mathparagraph }$ & $ +22.09\%  $

 & $\bf{0.4713}^{\dagger \ddagger \mathsection \mathparagraph }$ & $ +17.53\%  $

& 32,420/52,090/15,490

\\ \hline \hline
\end{tabular}

\end{table*}

\textbf{Baselines:} Three baselines from previous research are compared: \texttt{Frequency}, \texttt{PageRank}, and \texttt{LeToR}.

\texttt{Frequency}~\cite{dunietz2014new} estimates the salience of an entity by its term frequency. It is a straightforward but effective baseline in many related tasks. IDF is not as effective in entity-based text representations~\cite{ESR,raviv2016document}, so we used only frequency counts.

\texttt{PageRank}~\cite{dunietz2014new} estimates the salience score of an entity using its PageRank score~\cite{blanco2012graph}. We conduct a supervised PageRank on a fully connected graph. The nodes are the entities in the document. The edges are the embedding similarities of the connected nodes. The entity embeddings are configured and learned in the same manner as \texttt{KESM}. 
Similar to previous work~\cite{dunietz2014new}, \texttt{PageRank} is not as effective in the salience task. The results reported are from the best setup we found: a one-step random walk linearly combined with \texttt{Frequency}. 

\texttt{LeToR}~\cite{dunietz2014new} is a feature-based learning to rank (entity) model. It is trained using the same pairwise loss with \texttt{KESM}, which we found more effective than the pointwise loss used in prior research~\cite{dunietz2014new}. 

We re-implemented the features used by Dunietz and Gillick~\cite{dunietz2014new}. 
As listed in Table~\ref{tab:feature}, the features are extracted by various linguistic and semantic techniques including entity linking, dependency parsing, named entity recognition, and entity coreference resolution. Besides the standard Frequency count, the Head Word Count considers syntactic signals when counting entities; the Coreference Count considers all mentions that refer to an entity as its appearances when counting frequency.

The entity embeddings are trained on the same corpus using Google's Word2vec toolkit~\cite{word2vec}. Entity linking is done by TagMe; all entities are kept~\cite{raviv2016document, ESR}.
Other linguistic and semantic preprocessing are done by the Stanford CoreNLP toolkit~\cite{StanfordNLP}.

Compared to Dunietz and Gillick~\cite{dunietz2014new}, we do not include the headline feature because it uses information from the expert-written summary and does not improve the performance much anyway; we also replace the head-lex feature with Embedding Vote which has similar effectiveness but is more efficient.

\textbf{Evaluation Metrics:}
We use the ranking-focused evaluation metrics: Precision@\{1, 5\} and Recall@\{1, 5\}.
These metrics circumvent the problem of selecting a cutoff threshold for each individual document in classification evaluation metrics~\cite{dunietz2014new}.
Statistical significances are tested by permutation test with $p<0.05$.

\textbf{Implementation Details:} The hyper-parameters of \texttt{KESM} are configured following popular choices or previous research.
The dimension of entity embeddings, word embeddings, and CNN filters are all set to 128.
The kernel pooling layers use the same pre-defined kernels as in previous research~\cite{K-NRM}: 
one exact match kernel ($\mu=1, \sigma=1e-3$) and ten soft match kernels equally splitting the cosine similarity range $[-1, 1]$ ($\mu \in\{-0.9, -0.7,...,0.9\}$ and $\sigma=0.1$).
The length of the CNN used to encode entity description is set to 3 which is tri-gram.
The entity descriptions are fetched from Freebase. The first 20 words (the gloss sentence) of the description are used.
The words or entities that appear less than 2 times in the training corpus are replaced by ``Unk\_word'' or ``Unk\_entity''.

The parameters include the embeddings $V$, the CNN weights $W^c$, the projection weights $W^p$, and the kernel weights $W^s, b^s$. They are learned end-to-end using Adam optimizer, size 64 mini-batching, and early-stopping on the development split. 
$V$ is initialized by the skip-gram embeddings of words and entities jointly trained on the training corpora, which takes several hours~\cite{xiong2017duet}.
With our PyTorch implementation, \texttt{KESM} usually only needs one pass on the training data and converges within several hours on a typical GPU. 
In comparison, \texttt{LeToR} takes days to extract its features because parsing and coreference are costly.

\section{Salience Evaluation Results}
This section first presents the overall evaluation results for the entity salience task. Then it analyzes the advantages of modeling salience over counting frequency. 

\subsection{Entity Salience Performance}

Table~\ref{tab:overall_salience} shows the experimental results for the entity salience task. 
\texttt{Frequency} provides reasonable estimates of entity salience. 
The most frequent entity is often salient to the document; the Precision@1 is rather high, especially on the New York Times corpus. 
\texttt{PageRank} barely improves \texttt{Frequency}, although its embeddings are trained by the salience labels. 
\texttt{LeToR}, on the other hand, significantly improves both Precision and Recall of \texttt{Frequency}~\cite{dunietz2014new}, which is expected as it has much richer features from various sources.

\texttt{KESM} outperforms all baselines significantly. Its improvements over \texttt{LeToR} are more than $10\%$ on both datasets with only one exception: Precision@1 on New York Times. 
The improvements are also robust: About twice as many documents are improved (Win) than hurt (Loss). 

We also conducted ablation studies on the source of evidence in \texttt{KESM}.
Those marked with (E) include the entity kernels; those with (W) include word kernels; those with (K) enrich the entity embeddings with description embeddings. 
All variants include the entity kernels (E); otherwise the performances significantly dropped in our experiments.

\begin{figure*}[t]
\centering
\begin{subfigure}{0.47\textwidth}
\includegraphics[width=\textwidth]{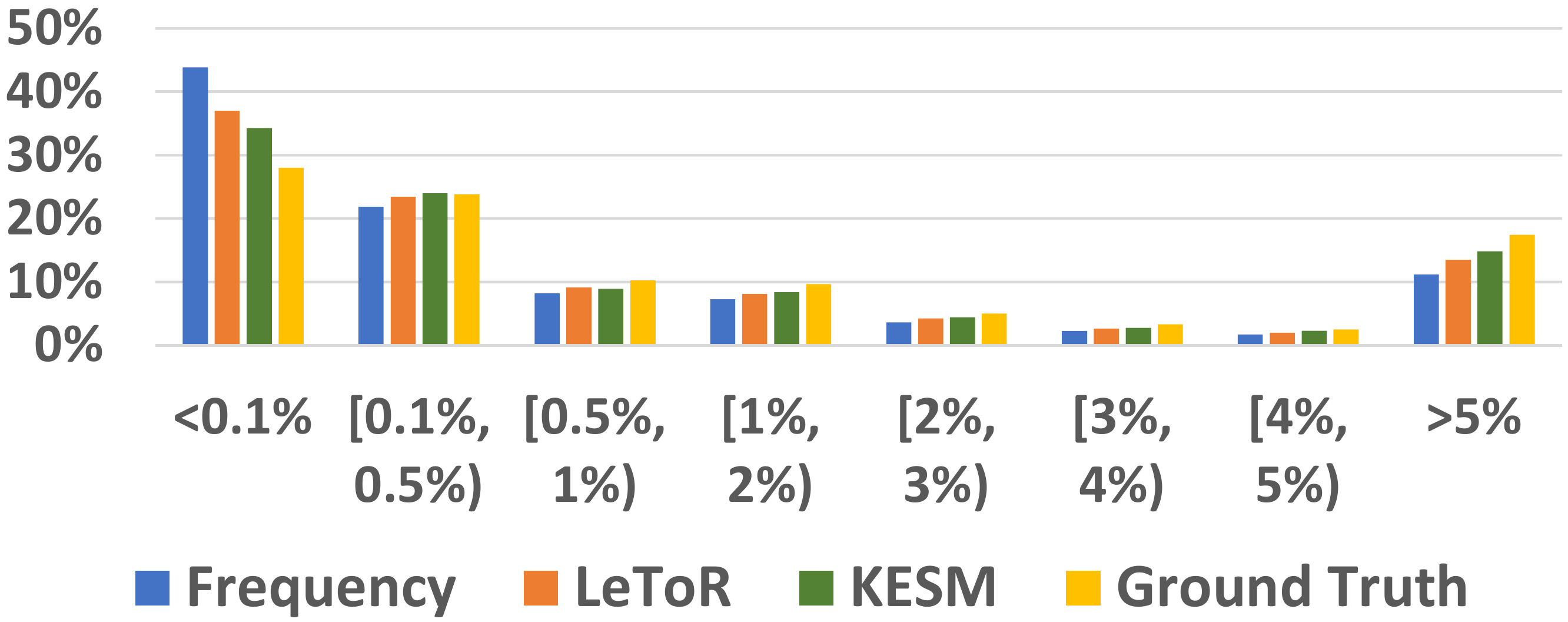}
\caption{New York Times \label{fig:nyt_dist}}
\end{subfigure}
\begin{subfigure}{0.47\textwidth}
\includegraphics[width=\textwidth]{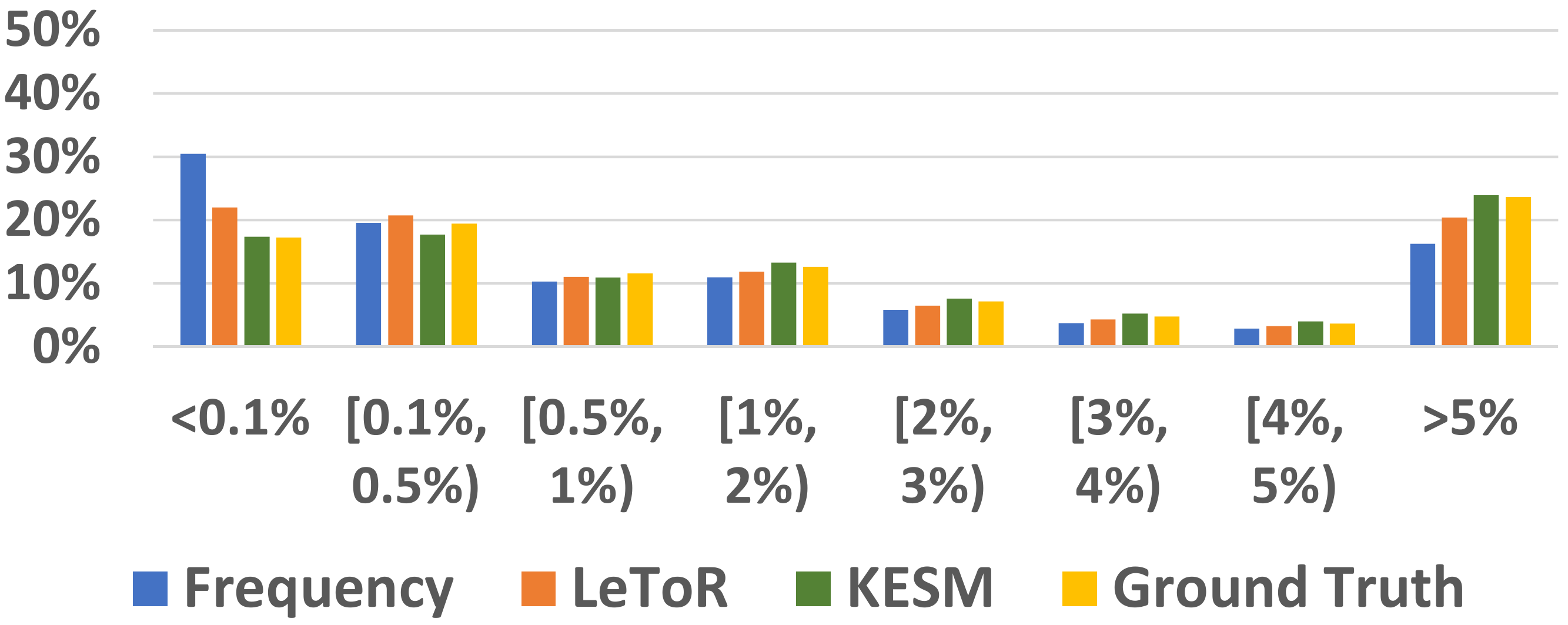}
\caption{Semantic Scholar \label{fig:s2_dist}}
\end{subfigure}
\caption{
The distribution of salient entities predicted by different models. The entities are binned by their frequencies in testing data. The bins are ordered from most frequent (Top 0.1\%) to less frequent (right).
The x-axes mark the percentile range of each group. The y-axes are the fraction of salient entities in each bin.
The histograms are ordered the same as the legends.
\label{fig:e_dist}
}
\end{figure*}

\texttt{KESM} performs better than all of its variants, showing that all three sources contributed.
Individually, \texttt{KESM (E)} outperforms all baselines. Compared to \texttt{PageRank}, the only difference is that \texttt{KESM (E)} uses kernels to model the interactions which are much more powerful than the raw embedding similarities used in \texttt{PageRank}~\cite{K-NRM}.
\texttt{KESM (EW)} always significantly outperforms \texttt{KESM (E)}.
The interaction between an entity and document words conveys useful information, the distributed representations make them easily comparable, and the kernels model the word-entity interactions effectively.
Knowledge enrichment (K) provides mixed results. A possible reason is that the training data is large enough to train good entity embeddings. Nevertheless, we find that adding the external knowledge makes the model stable and converged faster.

\subsection{Modeling Salience VS. Counting Frequency}
\label{sec:analysis}              
This experiment provides two analyses that study the advantage of \texttt{KESM} over counting frequency.

\textbf{Ability to Model Tail Entities.}
The first advantage of \texttt{KESM} is that it is able to model the salience of less frequent (tail) entities.
To demonstrate this effect, Figure~\ref{fig:e_dist} illustrates the distribution of predicted-salient entities in different frequency ranges.
The entities with top k highest predicted scores are predicted-salient, while k is the number of salient entities in the ground truth.

In both datasets, the frequency-based methods are highly biased towards the head entities:
The top $0.1\%$ most popular entities receive almost two-times more salience predictions from \texttt{Frequency} than in ground truth.
This is an intrinsic bias of frequency-based methods which not only limits their effectiveness but also attractiveness---less unexpected entities are selected.

In comparison, the distributions of \texttt{KESM} are much closer to the ground truth. \texttt{KESM} does a better job in modeling tail entities because it estimates salience not only by frequency but also by modeling the \emph{interactions} between entities and words. A tail entity can be estimated salient if many other entities and words in the document are closely related to it. For example, there are many entities and words describing various aspects of an entity in its Wikipedia page; the entities and words on a personal homepage are probably related to the person. These entities and words can `vote up' the title entity or the person because they are strongly connected to it/her. 
The ability to model such interactions with distributed representations and kernels is the main source of \texttt{KESM}'s text understanding capability.

\begin{figure}[t]
\centering
\begin{subfigure}{0.235\textwidth}
\includegraphics[width=\textwidth]{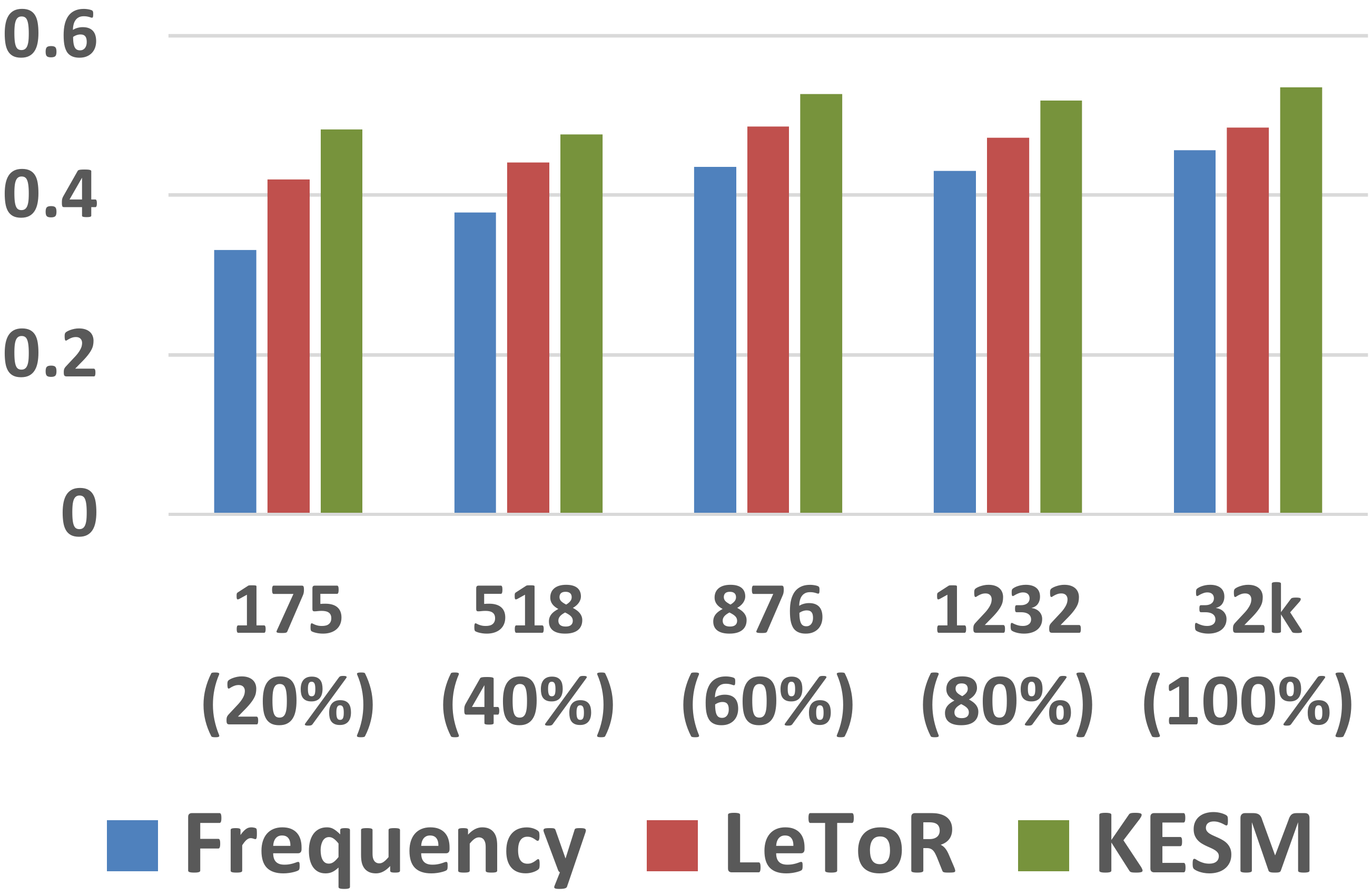}
\caption{New York Times \label{fig:nyt_doclen}}
\end{subfigure}
\begin{subfigure}{0.235\textwidth}
\includegraphics[width=\textwidth]{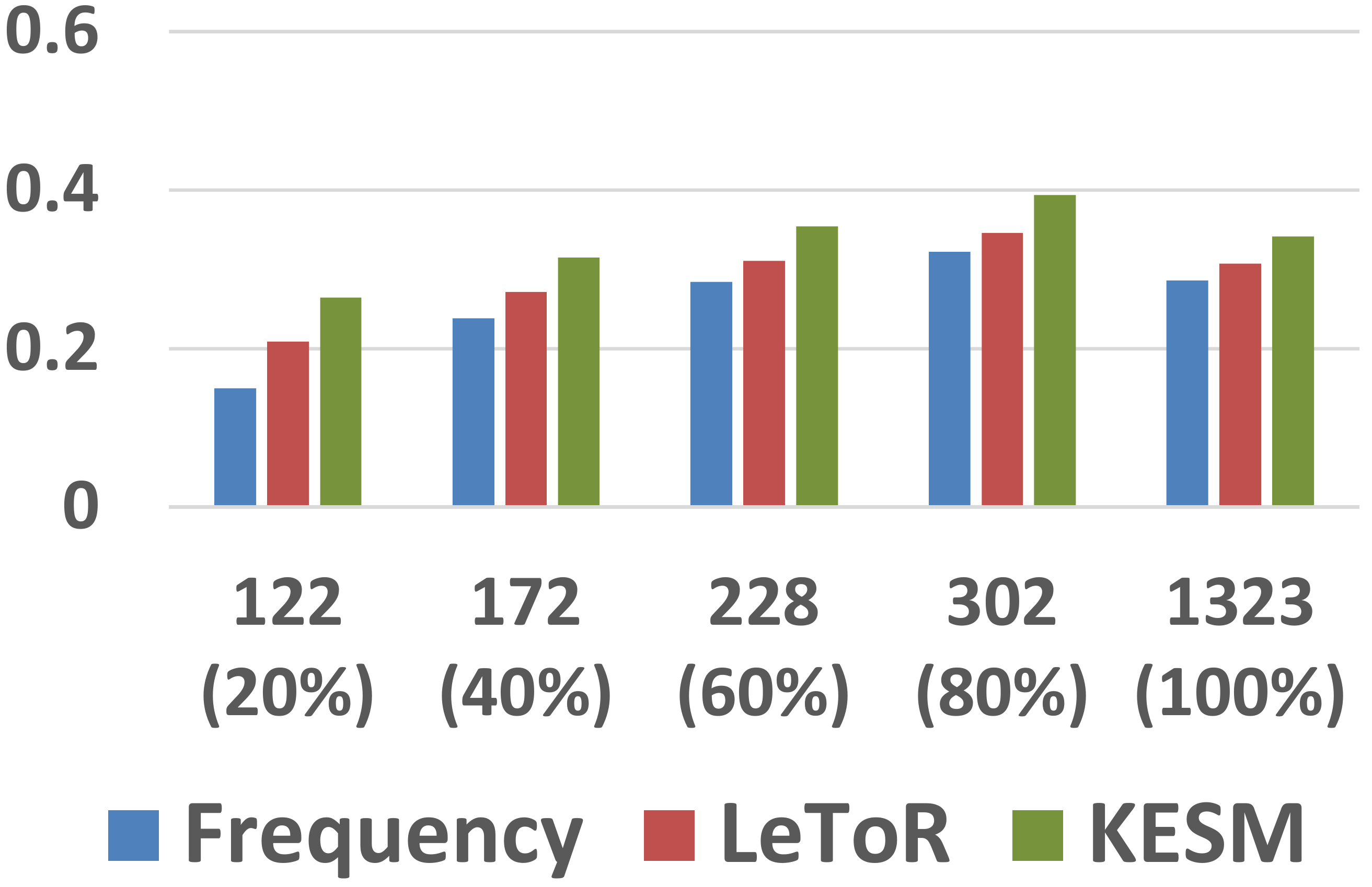}
\caption{Semantic Scholar \label{fig:s2_doclen}}
\end{subfigure}

\caption{Performances on documents with varying lengths (number of words).
The x-axes are the maximum length of the documents and the percentile of each group. 
The y-axes mark the performances on Precision@5.
The histograms are ordered the same as the legends.
\label{fig:doclen}
}
\end{figure}

\textbf{Reliable on Short Documents.}
The second advantage of \texttt{KESM} is its reliability on short texts.
To demonstrate it, we analyzed the performances of models on documents of varying lengths.
Figure~\ref{fig:doclen} groups the testing documents into five bins by their lengths (number of words), ordered from short (left) to long (right). Their upper bounds and percentiles are marked on the x-axes. The Precision@5 of corresponding methods are marked on the y-axes.

Both \texttt{Frequency} and \texttt{LeToR} (whose features are also mostly frequency-based) are less reliable on shorter documents.
The advantages of \texttt{KESM} are more significant when documents are shorter, 
while even in the longest bins where documents have thousands of words, \texttt{KESM} still outperforms \texttt{Frequency} and \texttt{LeToR}. 
Solely counting frequency is not sufficient to understand documents. 
The interactions between words and entities provide richer evidence and help \texttt{KESM} perform more reliably on shorter documents.

\begin{table*}[th]
\centering
\caption{
Ad hoc search accuracy of \texttt{KESM} when used as ranking features in learning to rank.
Relative performances over \texttt{IRFusion} are shown in the percentages. W/T/L are the number of queries a method improves, does not change, or hurts, compared with \texttt{IRFusion}. $\dagger$, $\ddagger$, $\mathsection$, and $\mathparagraph$ mark the statistically significant improvements over \texttt{BOE}$^\dagger$, \texttt{IRFusion}$^\ddagger$, \texttt{ESR}$^\mathsection$, and \texttt{ESR+IRFusion}$^\mathparagraph$. \texttt{BOW} is the base retrieval model, which is SDM in ClueWeb09-B and language model in ClueWeb12-B13.
\label{tab:overall_ranking}
}

\begin{tabular}{l|lr|lr|c||lr|lr|c}
\hline \hline
 & \multicolumn{5}{c||}{\bf{ClueWeb09-B}}
 & \multicolumn{5}{c}{\bf{ClueWeb12-B13}}
 \\ \hline 
\multicolumn{1}{c|}{\bf{Method}} &
\multicolumn{2}{c|}{\bf{NDCG@20}} &
\multicolumn{2}{c|}{\bf{ERR@20}} &
\bf{W/T/L} &
\multicolumn{2}{c|}{\bf{NDCG@20}} &
\multicolumn{2}{c|}{\bf{ERR@20}} &
\bf{W/T/L}
\\ \hline

\texttt{BOW}
 & ${0.2496}$ & $ -5.26\%  $

 & ${0.1387}$ & $ -10.20\%  $

& 62/38/100
 & ${0.1060}$ & $ -12.02\%  $

 & ${0.0863}$ & $ -6.67\%  $

& 35/22/43

\\
\texttt{BOE}
 & ${0.2294}$ & $ -12.94\%  $

 & ${0.1488}$ & $ -3.63\%  $

& 74/25/101

 & ${0.1173}$ & $ -2.64\%  $

 & ${0.0950}$ & $ +2.83\%  $

& 44/19/37

\\
\hline

\texttt{IRFusion}
 & $0.2635$ & --  & $0.1544$ & --  & --/--/--
 & $0.1205$ & --  & $0.0924$ & --  & --/--/--\\

\texttt{ESR}
 & ${0.2695}^{\dagger }$ & $ +2.30\%  $

 & ${0.1607}$ & $ +4.06\%  $

& 80/39/81
 & ${0.1166}$ & $ -3.22\%  $

 & ${0.0898}$ & $ -2.81\%  $

& 30/23/47

\\ \hline
\texttt{KESM}
 & ${0.2799}^{\dagger }$ & $ +6.24\%  $

 & ${0.1663}$ & $ +7.68\%  $

& 85/35/80
 & ${0.1301}^{\dagger \mathsection }$ & $ +7.92\%  $

 & $\bf{0.1103}^{\ddagger \mathsection \mathparagraph }$ & $ +19.35\%  $

& 43/25/32
\\
\hline
\hline
\texttt{ESR+IRFusion}
 & ${0.2791}^{\dagger \ddagger }$ & $ +5.92\%  $

 & ${0.1613}$ & $ +4.46\%  $

& 91/34/75

 & ${0.1281}$ & $ +6.30\%  $

 & ${0.0951}$ & $ +2.87\%  $

& 45/24/31

\\
\texttt{KESM+IRFusion}
 & $\bf{0.2993}^{\dagger \ddagger \mathsection \mathparagraph }$ & $ +13.58\%  $

 & $\bf{0.1797}^{\dagger \ddagger \mathsection \mathparagraph }$ & $ +16.38\%  $

& 98/35/67
& $\bf{0.1308}^{\dagger \mathsection }$ & $ +8.52\%  $

 & ${0.1079}^{\ddagger \mathsection \mathparagraph }$ & $ +16.77\%  $

& 43/23/34

\\
\hline \hline

\end{tabular}
\end{table*}

\begin{table*}[t]
\centering
\caption{
Ranking performances of \texttt{IRFusion}, \texttt{ESR}, and \texttt{KESM} with title or body field individually.
Relative performances (percentages) and Win/Tie/Loss are calculated by comparing with \texttt{IRFusion} on the same field.
$\dagger$ and $\ddagger$ mark the statistically significant improvements over \texttt{IRFusion}$^\dagger$ and \texttt{ESR}$^\ddagger$, also on the same field.
\label{tab:field_ranking}
}

\begin{tabular}{l|lr|lr|c||lr|lr|c}
\hline \hline
 & \multicolumn{5}{c||}{\bf{ClueWeb09-B}}
 & \multicolumn{5}{c}{\bf{ClueWeb12-B13}}
 \\ \hline 
\multicolumn{1}{c|}{\bf{Method}} &
\multicolumn{2}{c|}{\bf{NDCG@20}} &
\multicolumn{2}{c|}{\bf{ERR@20}} &
\bf{W/T/L} &
\multicolumn{2}{c|}{\bf{NDCG@20}} &
\multicolumn{2}{c|}{\bf{ERR@20}} &
\bf{W/T/L}
\\ \hline
\texttt{IRFusion-Title}
  & ${0.2584}$ & $ -3.51\%  $

 & ${0.1460}$ & $ -5.16\%  $

& 83/48/69

 & ${0.1187}$ & $ +6.23\%  $

 & ${0.0894}$ & $ +3.14\%  $

& 41/23/36

 \\
\texttt{ESR-Title}
 & $0.2678$ & --  & $0.1540$ & --  & --/--/--
 & $0.1117$ & --  & $0.0867$ & --  & --/--/--

\\ \hline
\texttt{KESM-Title}
 & $\bf{0.2780}^{\dagger }$ & $ +3.81\%  $

 & $\bf{0.1719}^{\dagger \ddagger }$ & $ +11.64\%  $

& 91/46/63

 & $\bf{0.1199}$ & $ +7.36\%  $

 & $\bf{0.0923}$ & $ +6.42\%  $

& 35/28/37

\\ \hline \hline

\texttt{IRFusion-Body}
 & ${0.2550}$ & $ +0.48\%  $

 & ${0.1427}$ & $ -3.44\%  $

& 80/46/74

 & ${0.1115}$ & $ +4.61\%  $

 & ${0.0892}$ & $ -3.51\%  $

& 36/30/34
\\
\texttt{ESR-Body}
& $0.2538$ & --  & $0.1478$ & --  & --/--/--
& $0.1066$ & --  & $0.0924$ & --  & --/--/--
\\ \hline
\texttt{KESM-Body}
  & $\bf{0.2795}^{\dagger \ddagger }$ & $ +10.13\%  $

 & $\bf{0.1661}^{\dagger \ddagger }$ & $ +12.37\%  $

& 96/39/65

  & $\bf{0.1207}^{\ddagger }$ & $ +13.25\%  $

 & $\bf{0.1057}^{\dagger \ddagger }$ & $ +14.44\%  $

& 43/24/33
\\
\hline \hline

\end{tabular}
\end{table*}

\section{Experimental Methodology \\ 
for Ad Hoc Search}
\label{sec:adapt}
This section presents the experimental methodology for the ad hoc search task. It follows a popular setup in recent entity-oriented search research~\cite{xiong2017duet}\footnote{Available at \url{http://boston.lti.cs.cmu.edu/appendices/SIGIR2017\_word\_entity\_duet/}}.

\textbf{Datasets} are from the TREC Web Track ad hoc search tasks, a widely used search benchmark. It includes 200 queries for the ClueWeb09 corpus and 100 queries for the ClueWeb12 corpus. The `Category B' subsets of the two corpora and corresponding relevance judgments are used.

The ClueWeb09-B rankings re-ranked the top 100 documents retrieved by sequential dependency model (SDM) queries~\cite{metzler2005markov} with standard post-retrieval spam filtering~\cite{daltonentity}.
On ClueWeb12-B13, SDM queries are not better than unstructured queries, and spam filtering provides mixed results; thus, we used unstructured queries and no spam filtering on this dataset, as in prior research~\cite{xiong2017duet}.
All documents were parsed by Boilerpipe to title and body fields~\cite{boilerpipe}.
The query and document entities are from Freebase and were annotated by TagMe~\cite{TagMe}.  All entities are kept. It leads to high coverage and medium precision, the best setting found in prior research~\cite{Xiong2016BOE}.

\textbf{Evaluation Metrics} are NDCG@20 and ERR@20, official evaluation metrics of TREC Web Tracks. Statistical significances are tested by permutation test (randomization test) with $p<0.05$.

\textbf{Baselines:}
The goal of our experiments is to explore the usage of entity salience modeling in ad hoc search. 
To this purpose, our experiments focus on evaluating the effectiveness of \texttt{KESM}'s entity salience features in standard learning to rank; the proper baselines are the ranking features from word-based matches (\texttt{IRFusion}) and entity-based matches (\texttt{ESR}~\cite{ESR}).
Unsupervised retrieval with words (\texttt{BOW}) and entities (\texttt{BOE}) are also included.

\texttt{BOW} is the base retrieval model, which is SDM on ClueWeb09-B and Indri language model on ClueWeb12-B.

\texttt{BOE} is the frequency-based retrieval with bag-of-entities~\cite{xiong2017duet}. It uses TagMe annotations and exact-matches query and documents in the entity space.
It performs similarly to the entity language model~\cite{raviv2016document} as they use the same information.

\texttt{IRFusion} uses standard word-based IR features such as language model, BM25, and TFIDF, applied to body and title fields. It is obtained from previous research~\cite{xiong2017duet}.

\texttt{ESR} is the entity-based ranking features obtained from previous research~\cite{xiong2017duet}. 
It includes both exact and soft match signals in the entity space~\cite{ESR}. The differences with \texttt{KESM} are that in \texttt{ESR}, the query and documents are represented by frequency-based bag-of-entities~\cite{ESR} and the entity embeddings are pre-trained in the relation inference task~\cite{bordes2013translating}.

\textbf{Implementation Details:}
As discussed in Section~\ref{sec:rank}, the TREC benchmarks do not have sufficient relevance labels for effective end-to-end learning; we pre-trained the \texttt{KEE} and \texttt{KIM} of \texttt{KESM} using the New York Time corpus and used them to extract salience ranking features.
The entity salience features are combined by the same learning to rank model (RankSVM~\cite{ranksvm}) as used by \texttt{IRFusion} and \texttt{ESR}, with the same cross validation setup~\cite{xiong2017duet}.
Similar to \texttt{ESR}, the base retrieval score is included as a feature in \texttt{KESM}.
In addition, we also concatenate the features of \texttt{ESR} or \texttt{KESM} to \texttt{IRFusion} to evaluate their effectiveness when combined with word-based features. The resulting feature sets \texttt{ESR+IRFusion} and \texttt{KESM+IRFusion} were evaluated exactly the same as they were individually.

As a result, the comparisons of \texttt{KESM} with \texttt{LeToR} and \texttt{ESR} hold out all other factors and directly investigate the effectiveness of the salience ranking features in a widely used learning to rank model (RankSVM).
Given the current exploration stage of entity salience in information retrieval,
we believe this is more informative than mixing entity salience signals into more sophisticated ranking systems~\cite{EsdRank, xiong2017duet}, in which many other factors come into play.

\begin{table*}[th]
\caption{Examples from queries that \texttt{KESM} improved or hurt, compared to \texttt{ESR}. 
Documents are selected from those that ESR and KESM disagreed.
The descriptions are manually written to reflect the main topics of the documents. 
\label{tab:eg}
}
\begin{tabular}{c|c|l|l}
\hline \hline
\multicolumn{4}{c}{\textbf{Cases that KESM Improved}} \\ \hline

\textbf{Query} & \textbf{Query Entities}  & \textbf{ESR Preferred Document}  & \textbf{KESM Preferred Document} \\ \hline 

\multirow{2}{*}{ER TV Show} & ``ER (TV Series)" & clueweb09-enwp02-22-20096 & clueweb09-enwp00-55-07707  \\
& ``TV Program" & ``List of films in Wiki without article" & ``ER ( TV series ) - Wikipedia"  \\
\hline

\multirow{2}{*}{Wind Power
} &
\multirow{2}{*}{``Wind Power
''} 
& clueweb12-0200wb-66-32730 & clueweb12-0009wb-54-01932  \\

&  & ``Home solar power systems" & ``Wind energy | Alternative Energy HQ"  \\
\hline

Hurricane Irene  
 &
 ``Hurricane Irene"
& clueweb12-0705wb-49-04059 & clueweb12-0715wb-81-29281
 \\

Flooding in Manville NJ &  ``Flood";  ``Manville, NJ" & ``Disaster funding for Hurricane Irene" & ``Videos and news about Hurricane Irene"  \\
\hline

\hline \hline
\multicolumn{4}{c}{\textbf{Cases that KESM Hurt}} \\ \hline

\textbf{Query} & \textbf{Query Entities}  & \textbf{ESR Preferred Document}  & \textbf{KESM Preferred Document} \\ \hline 

\multirow{2}{*}{Fickle Creek Farm
} &  ``Malindi Fickle"  & clueweb09-en0003-97-27345
 & clueweb09-en0005-66-00576
  \\
& ``Stream"; ``Farm" & ``Hotels near Fickle Creak" 
& ``List of breading farms"  \\
\hline

\multirow{3}{*}{Illinois State Tax
} &
``Illinois"; 
& clueweb09-enwp01-67-20725
 & clueweb09-en0011-23-05274
  \\

&  ``State Government" 
& ``Sales taxes in the United & ``Retirement-related general  \\

& ``US Tax"
& States, Wikipedia''
& purpose taxes by State''
\\

\hline

 \multirow{2}{*}{Battles in the Civil War}
 &
 ``Battles"
& clueweb09-enwp03-20-07742
 & clueweb09-enwp01-30-04139
 \\

 &  ``Civil War" & ``List of American Civil War battles" & ``List of wars in the Muslim world"  \\
\hline
\end{tabular}

\end{table*}

\section{Search Evaluation Results}
This section presents the evaluation results and case study in the ad hoc search task.

\subsection{Overall Result}

Table~\ref{tab:overall_ranking} lists the ranking evaluation results.
The three supervised methods, \texttt{IRFusion}, \texttt{ESR}, and \texttt{KESM}, all use the exact same learning to rank model (RankSVM) and only differ in their features.
\texttt{ESR+IRFusion} and \texttt{KESM+IRFusion} concatenate the two feature groups and use RankSVM to combine them. 

On both ClueWeb09-B and ClueWeb12-B13, \texttt{KESM} features are more effective than \texttt{IRFusion} and \texttt{ESR} features. On ClueWeb12-B13, \texttt{KESM} individually outperforms other features significantly by $8-20\%$. On ClueWeb09-B, \texttt{KESM} provides more novel ranking signals; \texttt{KESM+IRFusion} significantly outperforms \texttt{ESR+IRFusion}. The fusion on ClueWeb12-B13 (\texttt{KESM+LeToR}) is not as successful perhaps because of the limited ranking labels on ClueWeb12-B13.

To better investigate the effectiveness of entity salience in search, we evaluated the features on individual document fields. Table~\ref{tab:field_ranking} shows the ranking accuracies of the three feature groups when only the title field (\texttt{Title}) or the body field (\texttt{Body}) is used.
As expected, \texttt{KESM} is more effective on the body field than on the title field:
Titles are less noisy and perhaps all title entities are salient---not much new information is provided by salience modeling;
on the other hand, body texts are longer and more complicated, providing more opportunities for better text understanding.

The salience ranking features also behave differently with \texttt{ESR} and \texttt{IRFusion}. As shown by the W/T/L ratios in Table~\ref{tab:overall_ranking} and Table~\ref{tab:field_ranking}, more than $70\%$ query rankings are changed by \texttt{KESM}.
The ranking evidence provided by \texttt{KESM} features is from the interactions of query entities with the entities and words in the candidate documents. This evidence is learned from the entity salience corpus and is hard to be described by traditional frequency-based features.

\subsection{Case Study}
\label{sec:case}

The last experiment provides case studies on how \texttt{KESM} transfers its text understanding ability to search, by comparing the rankings of \texttt{KESM-Body} with \texttt{ESR-Body}. Both \texttt{ESR} and \texttt{KESM} match query and documents in the entity space, but \texttt{ESR} uses frequency-based bag-of-entities to represent documents while \texttt{KESM} uses entity salience.
We picked the queries where \texttt{KESM-Body} improved or hurt compared to \texttt{ESR-Body} and manually examined the documents they disagreed. The examples are listed in Table~\ref{tab:eg}. 

The improvements from \texttt{KESM} are mainly from its ability to determine whether a candidate document emphasizes the query entities or just mentions the query terms. 
As shown in the top half of Table~\ref{tab:eg}, \texttt{KESM} promotes documents where the query entities are more salient: the Wikipedia page about the ER TV show, a homepage about wind power, and a news article about the hurricane. On the other hand, \texttt{ESR}'s frequency-based ranking might be confused by web pages that only partially talk about the query topic. It is hard for ESR to exclude those web pages because they also mention the query entities multiple times.

Many errors \texttt{KESM} made are due to the lack of text understanding on the query side. \texttt{KESM} focuses on modeling the salience of entities in the \emph{candidate documents} and its ranking model treats all query entities equally.
As shown in the lower half of Table~\ref{tab:eg}, the query entities may contain errors, for example, ``Malindi Fickle'', or general entities that blur the (perhaps implied) query intent, for example ``Civil War'', ``State government'', and ``US Tax'.
These query entities do not align well with the information needs and thus mislead \texttt{KESM}.
Modeling the entity salience in \emph{queries} is a different task which is more about understanding search intents. 
To address these error cases may require a deeper fusion of \texttt{KESM} in more sophisticated ranking systems that can handle noisy query entities~\cite{xiong2017duet,JointSem}.

\section{Conclusion}

This paper presents \texttt{KESM}, the Kernel Entity Salience Model that estimates the salience of entities in documents.
\texttt{KESM} represents entities and words with distributed representations,  models their interactions using kernels, and combines the kernel scores to estimate entity salience. The semantics of entities in the knowledge graph---their descriptions---are also incorporated to enrich entity embeddings. 
In the entity salience task, the whole model is trained end-to-end using automatically generated salience labels.

In addition to the entity salience task, \texttt{KESM} is also applied to ad hoc search and ranks documents by the salience of query entities in them.
It calculates the kernel scores of query entities in the document,  combines them to salience ranking features, and uses a ranking model to predict the query-document ranking score.
When ranking labels are scarce, the ranking features can be extracted by pre-trained distributed representations and kernels from the entity salience task and then used by standard learning to rank.
These ranking features convey \texttt{KESM}'s text understanding ability learned from entity salience labels to search.

Our experiments on two entity salience corpora, a news corpus (New York Times) and a scientific publication corpus (Semantic Scholar), demonstrate the effectiveness of \texttt{KESM} in the entity salience task.
Significant and robust improvements are observed over frequency and feature-based methods. 
Compared to those baselines, \texttt{KESM} is more robust on tail entities and shorter documents; its Kernel Interaction Model is more powerful than the raw embedding similarities in modeling term interactions.
Overall, \texttt{KESM} is a stronger model with a more powerful architecture.

Our experiments on ad hoc search were conducted on the TREC Web Track queries and two ClueWeb corpora. In both corpora, the salience features provided by \texttt{KESM} trained on the New York Times corpus outperform both word-based ranking features and frequency-based entity-oriented ranking features, despite differences between the salience task and the ranking task. The advantages of the salience features are more observed on the document bodies on which deeper text understanding is required.

Our case studies on the winning and losing queries of \texttt{KESM} illustrate the influences of the salience ranking features: they distinguish documents in which the query entities are the core topic from those where the query entities are only partial to their central ideas. 
Interestingly, this leads to both winning cases---better text understanding leads to more accurate search---and also losing cases: when the query entities do not align well with the underlying search intent, emphasizing them ends up misleading the document ranking.

We find it very encouraging that \texttt{KESM} successfully transfers the text understanding ability from entity salience estimation to search.
Estimating entity salience is a fine-grained text understanding task that focuses on the detailed interactions between entities and words. Previously it was uncommon for text processing techniques at this granularity to be as effective in information retrieval. Often shallower methods worked better for search. However, the fine-grained text understanding provided by \texttt{KESM}---the interaction and consistency between query entities with the document entities and words---actually improves the ranking accuracy.
We view this work as an encouraging step from ``search by matching'' to ``search with meanings''~\cite{bast2016semantic} and hope it will motivate more future explorations towards this direction.

\section{Acknowledgments}
This research was supported by National Science Foundation (NSF) grant IIS-1422676 and DARPA grant FA8750-12-2-0342 under the DEFT program. Any opinions, findings, and conclusions in this paper are the authors' and do not necessarily reflect the sponsors'.


\bibliographystyle{ACM-Reference-Format}
\normalsize
\bibliography{citation}
\end{document}